\documentclass[twocolumn,dvipsnames]{aastex62}
\usepackage{amsmath,amstext}
\usepackage{apjfonts}
\usepackage[figure,figure*]{hypcap}
\usepackage{ulem}
\usepackage{xspace}
\usepackage{xcolor}
\usepackage{lineno}
\usepackage{xfrac}
\usepackage{graphicx}
\usepackage{natbib}
\usepackage{multimedia}
\usepackage{amssymb}
\usepackage{amsmath}
\usepackage{array,booktabs}
\usepackage{mathptmx}
\usepackage{booktabs}
\usepackage{hyperref}
\usepackage{float}
\usepackage{todonotes}
\usepackage{multirow}
\newcommand{\mbh}{\,{M_{\rm BH}}}
\newcommand{\dmbh}{\,{\dot{M}_{\rm BH}}}

\newcommand{\erg}{\,{{\rm erg}}}

\newcommand{\msun}{\,{M_{\odot}}}

\newcommand{\tBH}{\,{{t_{\rm BH}}}}

\newcommand{\s}{\,{{\rm s}}}

\newcommand{\cm}{\,{{\rm cm}}}

\shorttitle{Spinning into the Gap: Direct-Horizon Collapse as the Origin of GW231123}

\shortauthors{Gottlieb et al.}

\begin{document}

\title{Spinning into the Gap: Direct-Horizon Collapse as the Origin of GW231123 from End-to-End GRMHD Simulations}

\author[0000-0003-3115-2456]{Ore Gottlieb}
\email{oregottlieb@gmail.com}
\affiliation{Center for Computational Astrophysics, Flatiron Institute, 162 5th Avenue, New York, NY 10010, USA}
\affil{Department of Physics and Columbia Astrophysics Laboratory, Columbia University, Pupin Hall, New York, NY 10027, USA}
\affil{Department of Physics and Kavli Institute for Astrophysics and Space Research, Massachusetts Institute of Technology, Cambridge, MA 02139, USA}

\author[0000-0002-4670-7509]{Brian D. Metzger}
\affiliation{Department of Physics and Columbia Astrophysics Laboratory, Columbia University, Pupin Hall, New York, NY 10027, USA}
\affiliation{Center for Computational Astrophysics, Flatiron Institute, 162 5th Avenue, New York, NY 10010, USA}

\author[0009-0005-2478-7631]{Danat Issa}
\affiliation{Center for Interdisciplinary Exploration \& Research in Astrophysics (CIERA), Physics and Astronomy, Northwestern University, Evanston, IL 60201, USA}

\author[0000-0002-6341-4484]{Sean E. Li}
\affiliation{Department of Physics and Columbia Astrophysics Laboratory, Columbia University, Pupin Hall, New York, NY 10027, USA}

\author[0000-0002-6718-9472]{Mathieu Renzo}
\affil{University of Arizona, Department of Astronomy \& Steward Observatory, 933 N. Cherry Ave., Tucson, AZ 85721, USA}

\author[0000-0001-8830-8672]{Maximiliano Isi}
\affiliation{Department of Astronomy and Columbia Astrophysics Laboratory, Columbia University, Pupin Hall, New York, NY 10027, USA}
\affiliation{Center for Computational Astrophysics, Flatiron Institute, 162 5th Avenue, New York, NY 10010, USA}

\begin{abstract}

GW231123, the most massive binary black hole (BH) merger observed to date, involves component BHs with masses inside the pair-instability mass gap and unusually high spins. This challenges standard formation channels such as classical stellar evolution and hierarchical mergers. However, stellar rotation and magnetic fields, which have not been systematically incorporated in prior models, can strongly influence the BH properties. We present the first self-consistent simulations tracking a massive, low-metallicity helium star from helium core burning through collapse, BH formation, and post-BH formation accretion using 3D general-relativistic magnetohydrodynamic (GRMHD) simulations. Starting from a $250\,M_\odot$ helium core, we show that collapse above the pair-instability mass gap, aided by rotation and magnetic fields, drives mass loss through disk winds and jet launching. This enables the formation of highly spinning BHs within the mass gap and reveals a BH spin--mass correlation. Strong magnetic fields extract angular momentum from the BH through magnetically driven outflows, which in turn suppress accretion, resulting in slowly spinning BHs within the mass gap. In contrast, stars with weak fields permit nearly complete collapse and spin-up of the BH to $ a\approx1$. We show that massive low-metallicity stars with moderate magnetic fields naturally produce BHs whose masses and spins match those inferred for GW231123, and are also consistent with those of GW190521. The outflows launched during collapse may impart a BH kick, which can induce spin--orbit misalignment and widen the post-collapse orbit, delaying the merger. These outflows could further drive short-lived, high-luminosity jets comparable to the most energetic $\gamma$-ray bursts, offering a potential observational signature of such events in the early universe.

\end{abstract}

\section{Introduction}\label{introduction}

On November 23, 2023, the LIGO-Virgo-KAGRA (LVK) collaboration detected gravitational waves (GWs) from the binary black hole (BBH) merger GW231123, which occurred at a luminosity distance of $ D_L = 2.2^{+1.9}_{-1.5}\,{\rm Gpc} $ \citep{LIGO2025}. This event represents the most massive BBH ever observed by the LVK. Both BHs exhibit extreme properties, with high component masses of $ M_1 = 137^{+22}_{-17}\,\msun $ and $M_2 = 103^{+20}_{-52}\,\msun $, and high dimensionless spins of $ \chi_1 = 0.9^{+0.10}_{-0.19} $ and $ \chi_2 = 0.8^{+0.20}_{-0.51} $. The data also show signs of significant spin--orbit misalignment \citep{LIGO2025} but potential spin-spin alignment \citep{Bartos2025}. The extreme masses and spins of GW231123 pose a significant challenge to black hole (BH) formation theories, from hierarchical mergers to massive binary
evolution.

In hierarchical merger channels (e.g., \citealt{Gerosa2017}), at least one BH originates from a previous BBH merger, allowing it to grow to the high mass observed in GW231123. However, these mergers typically impart large GW recoil kicks that can unbind the remnant from its host system, requiring dense environments with high escape velocities \citep[e.g.,][]{Stone2017,Antonini2019,Rodriguez2019}. While dynamical assembly scenarios can produce spin--orbit misalignment due to random pairing and isotropic spin orientations, they rarely yield high spin magnitudes. Merger remnants generally have random orientations, which average out in-plane components and lead to modest effective precession. This makes it difficult to explain the unusually high effective precessing spin measured in GW231123 of $\chi_p = 0.77^{+0.17}_{-0.19}$ \citep[but see][who showed that third-generation BHs may still exhibit high spins, albeit in a very limited sample; and \citealt{Kirouglu2025}, who showed that highly spinning, mass-gap BHs may form through a single coherent encounter between a BH and a star in dense star clusters]{Bamber2025}. The combination of the improbable spin configuration and the requirement for a deep gravitational potential well makes the hierarchical formation channel a statistically disfavored origin for GW231123 \citep[see][and references therein]{Stegmann2025}. Gas accretion by a BBH embedded in an AGN disk could, in principle, increase the BH spin. Although estimates calibrated to the bulk of the LVK population generally predict only modest spins \citep[e.g.,][]{Tagawa+20,Chen&Lin23,McKernan&Ford24}, hierarchical mergers within the disk may yield highly spinning BHs \citep[e.g.,][]{Delfavero2025,Ford2025}.

In isolated stellar binary evolution, the main difficulty lies in the
fact that at least the secondary, and possibly also the primary BH in
GW231123, falls within the pair-instability supernova (PISN) mass gap
\citep[][]{Renzo2024}, estimated to span BH masses from
$M_{\rm low} \approx 69^{+32}_{-18}\,\msun$ to
$M_{\rm up} \approx 139^{+30}_{-14}\,\msun$
\citep[e.g.,][]{Belczynski2016,Farmer2019,Woosley2021,Mehta2022,Farag2022}.
A BH within the PISN mass gap may form through induced premature stellar collapse, such as via the capture of a small BH, but this pathway is unlikely \citep{Baumgarte&Shapiro25}. An alternative mechanism involving dynamical stellar mergers, which have been proposed to form BH-producing stars within the PISN gap \citep[e.g.,][]{Spera2019,DiCarlo2020,Kremer2020,Renzo2020,costa:22,ballone:23}. However, the predicted BH masses in this scenario are typically $M \lesssim 100\,\msun$ \citep{vanSon2020}, and the natal spins are expected to be similarly low to those from ordinary core-collapse SNe, $a \ll 1$  \citep{Heger2003, Fuller&Ma2019}, both in tension with the properties of GW231123. By contrast, chemically homogeneous evolution (CHE) of massive binary stars can produce BBH mergers with high spins, though current models have thus far been restricted to masses below those observed in GW231123 \citep{deMink2016,Mandel2016,Marchant2016,dubuisson:20}.

In evolved helium cores with masses $ M_{\rm He} > M_{\rm up}$, heavy
nuclei undergo photodisintegration due to the extremely high core
temperatures, resulting in significant thermal energy losses that
render the explosion too weak to halt gravitational collapse of the
massive core, ultimately leading to a ``direct-horizon'' collapse,
forming a massive BH \citep{Bond1984,Fryer2001,Heger2002,Heger2003}.
Sufficiently massive stars to reach this final end may have been
common among the first generation of stars, which could form with very
high masses due to the lack of efficient cooling in the early universe
\citep{Bromm1999,Abel2000}. But even at finite metallicity they may be
possible \citep[e.g.,][]{schneider:18b}: the most massive WNh stars
observed in the LMC exceed present-day masses of
$\gtrsim 200\,M_{\odot}$ \citep{dekoter:97, crowther:10, crowther:16,
  brands:22}, and although at metallicity
$Z_{\rm LMC}\simeq Z_{\odot}/2$, winds will reduce their mass in or
below the pair-instability gap before their collapse
\citep{renzo:19:vfts682}, similar stars at lower $Z$ may have reduced
winds and reach the direct collapse fate.

\begin{figure*}[]
  \centering
  \includegraphics[width=1\textwidth,trim={0cm 0cm 0cm 0cm}]{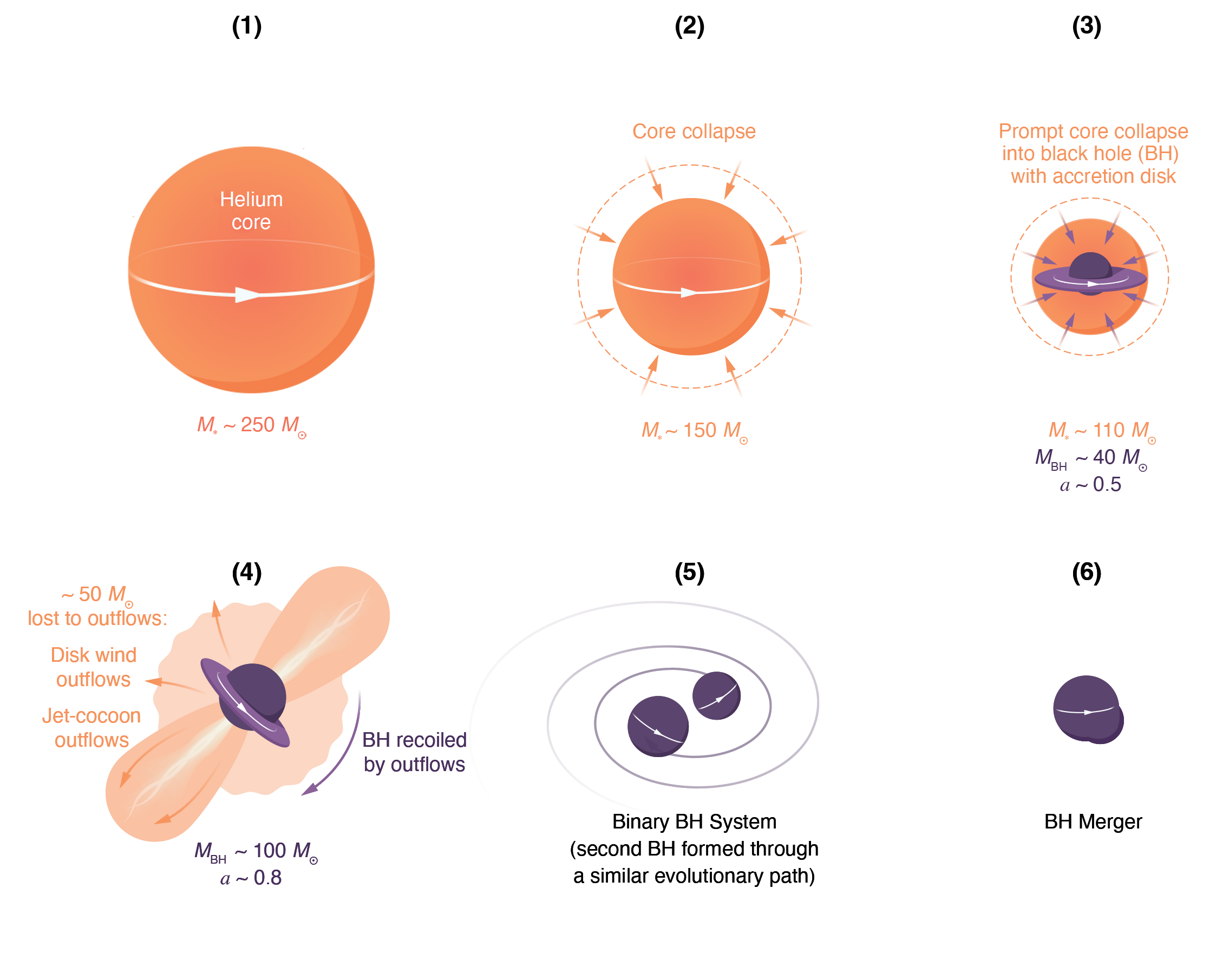}
  \caption{Outline of the evolutionary pathway leading to the GW231123 BBH merger, originating from a binary of low-metallicity massive stars that undergo direct-horizon collapse.
  {\bf (1)} The star initially contains a $ \sim 250\,\msun $ helium core, which is subject to mass loss through stellar winds and/or binary interaction.
  {\bf (2)} The helium core, now weighing $ \sim 150\,\msun $, has undergone fusion into carbon and oxygen and becomes unstable to pair production, triggering contraction.
  {\bf (3)} As the temperature rises, heavy nuclei begin to photodisintegrate in a runaway process that culminates in a direct-horizon collapse, forming a moderately spinning $ \sim 40\,\msun $ BH surrounded by an accretion disk formed as a result of the star's rotation.
  {\bf (4)} Magnetic fields threading the BH regulate its final spin and mass. They extract rotational energy, enabling the formation of BHs with a range of spin values, and drive mass loss through relativistic jets and disk winds, allowing BHs to form within the upper mass gap. A moderate seed magnetic field in the star yields a BH with $ M_{\rm BH} \approx 100\,\msun $ and $ a \approx 0.8 $. Turbulent accretion imparts a kick to the BH, potentially misaligning its spin axis relative to the binary orbit.
  {\bf (5)} A similar process occurs in the companion, but with weaker magnetic fields that extract less rotational energy from the BH and unbind less stellar mass. This results in the formation of a BH with $ M_{\rm BH} \approx 130\,\msun $ and $ a \approx 0.9 $. The independent evolution of each component leads to a BBH system with misaligned spins.
  {\bf (6)} The BBH merges after the inspiral driven by GW emission, yielding a $M_f \approx 225\,\msun$ remnant with spin $ a_f \approx 0.9$, consistent with GW231123.
}
  \label{fig:sketch}
\end{figure*}

In 
a direct collapse, the absence of a core bounce during prompt BH
formation precludes the generation of outflows, so BHs formed from
cores with $M_{\rm He} > M_{\rm up}$ are naively predicted to grow\footnote{Although core contraction releases
  gravitational binding energy, primarily in neutrinos, the associated
  rest-mass loss is only $ \lesssim \msun c^2 $, and therefore does
  not alter the conclusion that such BHs remain above the mass
  gap.} to
$M_{\rm BH} \approx M_{\rm He}$, placing them above the
pair-instability mass gap \citep{Renzo2020b}.\footnote{\cite{vink:21}
and more recently \citet{Tanikawa2025} proposed that if stellar evolution proceeds with inefficient convective overshooting and a dramatically reduced rate of the $^{12}$C($\alpha,\gamma$)$^{16}$O reaction, the growth of the helium core is suppressed and the upper mass gap is shifted to higher masses, enabling the formation of GW231123 BHs.}
However, rotation, which plays a crucial role in the evolution of low-metallicity massive stars \citep[e.g.,][]{Marigo2003,Ekstrom2008,Yoon2012,Popa2025}, may alter this outcome.\footnote{For example, extreme rapid rotation may shift the pair-instability mass gap to higher masses \citep{Croon2025}, thereby making it more difficult for the primary to lie above the gap.} Rapid rotation and magnetic fields can drive mass loss \emph{after} collapse through accretion disk winds or jet launching via the Blandford-Znajek mechanism \citep{Blandford&Znajek1977}, enabling the formation of a highly spinning BH within the mass gap \citep[e.g.,][]{Siegel+22}. Jets can also extract rotational energy from the BH via magnetic torques, potentially producing a wide range of final spins. In addition, magnetically driven explosions can also impart stochastic momentum to the BH, likely resulting in a potentially significant natal kick compared to unmagnetized direct collapse \citep[e.g.,][]{Burrows2025}, and inducing spin--orbit misalignment. If the kick velocity is not too large to unbind the binary, these effects make rapidly rotating stars a compelling candidate for the origin of the massive, high-spin BHs observed in GW231123. Figure~\ref{fig:sketch} outlines the evolutionary stages leading to the GW231123 BBH merger from a binary of two low-metallicity massive stars undergoing direct-horizon-collapse.

In this \emph{Letter}, we present the first end-to-end simulations of
direct-horizon-collapsing massive stellar cores. 
Our modeling begins with helium core evolution computed using
\textsc{mesa}, continues through gravitational collapse and BH
formation, and culminates in high-resolution 3D general-relativistic
magnetohydrodynamic (GRMHD) simulations of the post-BH formation
evolution. This enables us to determine the properties of the BH
remnant and to investigate how the progenitor’s rotation and magnetic
field influence the natal mass, spin, and kick of the newborn BH. We
show that for a moderately magnetized progenitor, the resulting BH
properties are consistent with those inferred for GW231123, including
its high mass, rapid spin, and likely significant recoil velocity. The structure of the paper is as follows. In
\S\ref{sec:prebh}, we describe the stellar evolution leading to
collapse and BH formation. In \S\ref{sec:postbh}, we outline the
numerical setup of the post-BH formation GRMHD simulations. In
\S\ref{sec:evolution}, we present the simulation results, and in
\S\ref{sec:bh}, we discuss the resulting BH mass, spin and kick. We
summarize our findings and assess their compatibility with GW231123 in
\S\ref{sec:summary}.

\section{Stellar Evolution to Black Hole Formation}\label{sec:prebh}

We employ a helium core model\footnote{Publicly available at \url{https://zenodo.org/records/3406357}.} with an initial mass of $ M_{\rm He} =
250.75\,\msun $ from \citet{Renzo2020b}. The model is initialized as a bare helium core with metallicity $ Z = 0.001 $, so the hydrogen-rich envelope is excluded and all wind-driven mass loss is helium-rich. Mass-loss rates is implemented through stellar winds \citep[see e.g.,][]{Vink2005}, with rates prescribed by \citet{Hamann1995,Hamann1998} reduced by a factor of 10 \citep[e.g.,][]{brott:11}, and can be further enhanced through binary interactions \citep{Delgado1981}, reducing the core mass to $ M_{\rm He} \approx 149\,\msun $ by helium core depletion, at the onset of core collapse.

The progenitor evolution employs a reduced nuclear reaction network that is not designed to capture dynamics dominated by weak interactions \citep[e.g.,][]{Farmer2016,Renzo2024b}, nor the detailed core bounce and shock propagation in the innermost core. In the mass range considered here, however, collapse is initiated by pair instability, and the ensuing photodisintegration of heavy nuclei absorbs nearly all of the gravitational binding energy released during the pair-instability-driven contraction. As a result, although weak reactions act to deleptonize the gas, they are not the primary driver of the dynamics, which are instead governed by BH accretion of plasma and the associated feedback from outflows.

To model the collapse through to BH formation, we neglect the low-density outer part of the core and evolve the inner $ M_{\rm He} \approx 140\,\msun $ of the star, enclosed within a radius of $ R_{\rm He} \approx 5.6\times 10^9\,{\rm cm} $, using the 1D general-relativistic hydrodynamic code \textsc{gr1d}\footnote{Available at \url{https://github.com/evanoconnor/GR1D}.} \citep{OConnor2010}. \textsc{gr1d} solves the equations of non-rotating general relativistic hydrodynamics coupled to a microphysical, finite-temperature equation of state \citep[LS220;][] {Lattimer_LS220}. While LS220 may be in tension with nuclear experimental constraints \citep{Tews2017}, it remains widely used due to its consistency with multimessenger constraints from GW170817 \citep[e.g.,][]{Radice2018}. For equation of states consistent with GW170817, the dynamics, remnant properties, and ejecta are broadly similar, with differences mainly in quantitative details of the neutron star. Our results should therefore be robust to the specific choice of equation of state within this allowed set.

\begin{figure}[]
  \centering
  \includegraphics[width=0.5\textwidth]{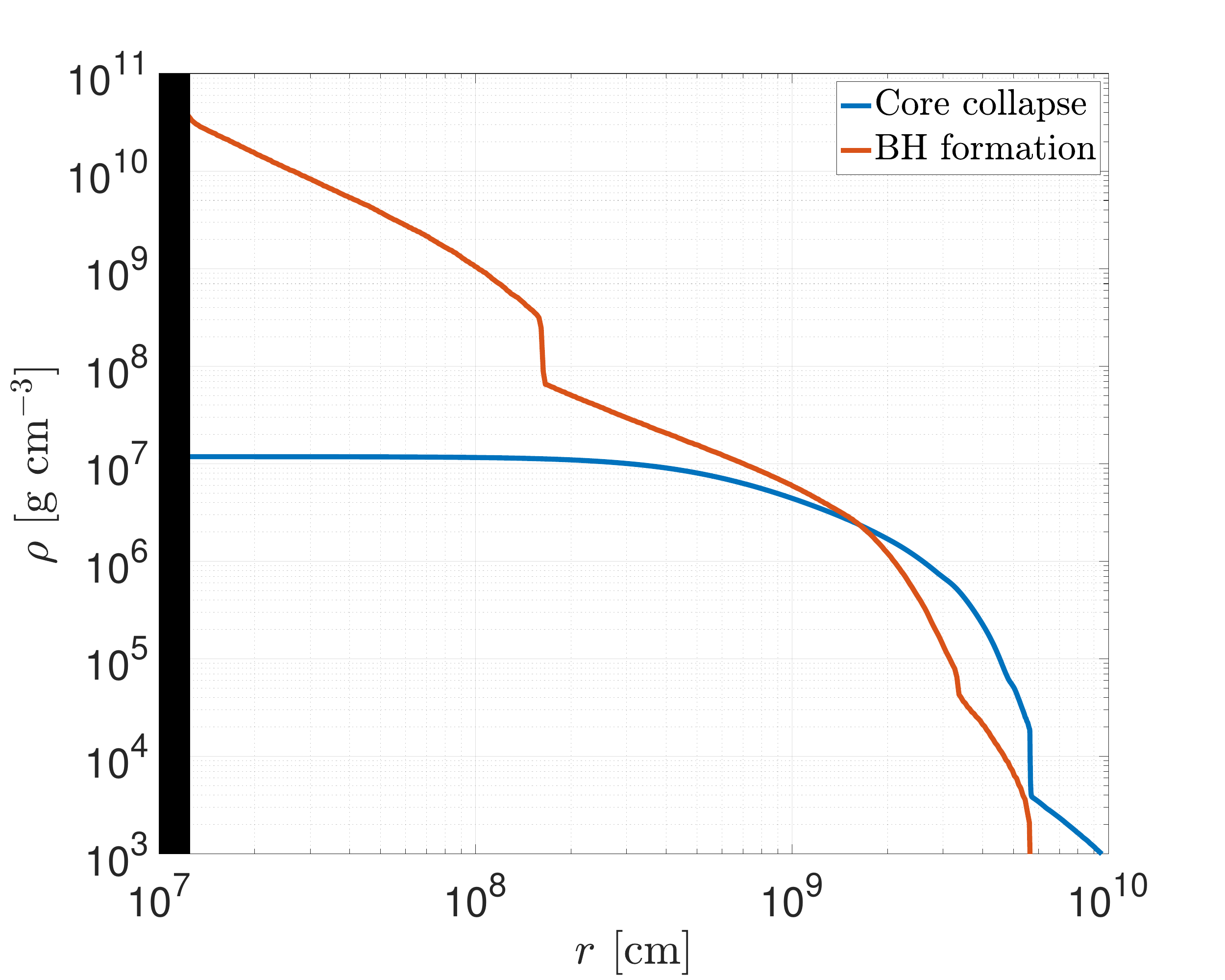}
  \caption{Evolution of the mass density profile from the onset of core collapse (blue) to $ \mbh \approx 40\,\msun $ BH formation at $ \tBH = 4.29\,\s $ after core collapse (red). The black area indicates the Schwarzschild radius of the BH, $ r_{\rm s} = 2G\mbh/c^2 \approx 1.2\times 10^7\,\cm $.
  }
  \label{fig:density}
\end{figure}

We choose a 1D grid with a total of 600 radial cells, which is uniform in the inner region ($r \leq 20\,\rm{km}$) and is logarithmically spaced outside of it. The edge of the star is set by a density threshold $\rho = 2\times10^3\,\rm{g\,cm^{-3}}$, which is close to the lowest density value in the tabulated equation of state. We consider a BH to form once the central density exceeds $\rho = 2\times10^{14}\,{\rm g\,cm^{-3}}$ or the central lapse falls below $\sim10^{-2}$; in our simulation, these criteria are satisfied at $t_{\rm BH} = 4.29\,\mathrm{s}$ after core collapse, when a $\mbh \approx 40\,\msun$ BH forms and \textsc{gr1d} terminates. Figure~\ref{fig:density} depicts the evolution of the mass density profile from the onset of collapse (blue) to the moment of BH formation (red), with the black area delineating the Schwarzschild BH horizon.

\section{Post-Black Hole Formation Modeling}\label{sec:postbh}

To model the post-BH formation phase (see Fig.~\ref{fig:sketch}), we
map the mass density and radial velocity profiles of the star at the
time of BH formation from \textsc{gr1d} into the 3D GRMHD code
\textsc{h-amr} \citep{Liska2022}. \textsc{h-amr} is a GPU-accelerated
code that enables high-resolution simulations of the system from BH formation through the accretion or ejection of all fluid elements in the collapsing star. However, \textsc{h-amr} does not evolve the spacetime metric -- the BH mass and spin remain fixed during the simulation. Likewise, \textsc{h-amr} does not include any Newtonian approximation for self-gravity, and may therefore slightly underestimate the accretion rate onto the BH. Therefore, in \S\ref{sec:bh} we estimate the final BH mass and spin in post-processing by computing the total mass and angular momentum accreted onto the BH. However, we caution that a fully self-consistent evolution of the BH mass and spin during the simulation, accounting for feedback processes, is required for accurate results.

Although the evolution prior to BH formation has included neither magnetic fields nor rotation, both are likely to play an important role in the collapse of low-metallicity massive stars \citep[e.g.,][]{Fryer2001,Croon2025,Popa2025}. However, the rotational profiles of these stars remain highly uncertain \citep[e.g.,][]{Heger2003}. For example, as the convective core contracts during stellar evolution, angular momentum is expected to be transported outward and partially lost, making the rotational structure at the onset of collapse difficult to predict. To incorporate rotation in our setup, we adopt a rotation profile motivated by lower-mass progenitor models. Specifically, we fit an average profile to the angular momentum distributions of 125 massive stars ($ M_\star \approx 30\,\msun $ at core collapse) with metallicity $Z = 0.001$, evolved to core collapse using \textsc{mesa} \citep{Paxton2011,Paxton2013,Paxton2015,Paxton2018,Paxton2019,Jermyn2023}. These stars begin with initial rigid rotation corresponding to $\omega \gtrsim 0.5\,\omega_{\rm crit}$, where $\omega_{\rm crit}$ is the surface critical rotation rate that includes radiative forces, ensuring rotationally induced CHE. Their rotational mixing follows \citet{Heger2000}, including angular momentum transport via Tayler-Spruit dynamo \citep{Spruit2002}. An example profile is shown in \citet{Gottlieb2024b}; the full sample will be presented in a forthcoming publication. As noted in \citet{Gottlieb2024b}, adopting a more efficient angular momentum transport mechanism \citep[e.g.,][]{Fuller2019, fuller:22} results in too little rotation for the formation of an accretion disk during the collapse.

The fit to the angular momentum profiles of CHE models consists of
three power-law segments, corresponding to the burning core, the
radiative envelope, and the convective outer envelope. 
We focus on the two inner regions for the specific
angular momentum profile, given by
\begin{equation}\label{eq:j}
\ell(r,\theta) =
\begin{cases}
0.55\,r^2\,{\rm sin}^2\theta\,\s^{-1} & r < r_c\\
&\\
0.55\,r_c^2\,{\rm sin}^2\theta\,\s^{-1} & r < r_c < R_{\rm He} \\
\end{cases} \,,
\end{equation}
where $ r_c \approx 4\times 10^8\,\cm $ is chosen by the constant density radius at the onset of collapse (see Fig.~\ref{fig:density}). This profile arises as the convective core contracts and transports angular momentum outward: the inner core remains in near-rigid rotation, while the outer part retains a roughly constant specific angular momentum. The newly formed BH is born with a dimensionless spin parameter \citep{Bardeen1970}
\begin{equation}\label{eq:a0}
    a = \frac{c}{G\mbh^2}\int_0^{\mbh} \ell(m)\mathrm{d}m\approx 0.5\,,
\end{equation}
where $ \ell(m) $ is the specific angular momentum of mass coordinate $ m $. Based on Equation~\eqref{eq:a0}, we adopt $ a = 0.5 $ as the BH spin in our simulation. Since the enclosed mass coordinate $ \mbh $ lies at $ r > r_c $, Equation~\eqref{eq:j} implies that all infalling plasma after BH formation carries the same specific angular momentum.

We initialize the magnetic field assuming it is aligned with the stellar rotation axis and follows a radial profile that falls off slower/like/faster than a dipole field, thereby determining the magnetic field strength of plasma accreting onto the disk from large radii. To ensure that the magnetic field smoothly vanishes at the stellar boundary, we add a tapering term such that the vector potential takes the form
\begin{equation}\label{eq:A}
    A=A_\phi(r<R_{\rm He},\theta) = \mu\,\frac{\rm sin\theta}{r}\, \left(\frac{r}{r_g}\right)^n\cdot \bigg(\frac{1}{r} - \frac{1}{R_{\rm He}}\bigg)~,
\end{equation}
where $ \mu $ is the magnetic moment. The model names, along with the magnetic field strength, $ \mu $, and power-law index, $ n $, are listed in Table~\ref{tab:models}.

\begin{table}[]
    \setlength{\tabcolsep}{8.8pt}
    \centering
    \renewcommand{\arraystretch}{1.2}
    \begin{tabular}{| c | c c | }

            \hline
        Model & $ {\rm log_{10}}B~[{\rm G}]~(r=100\,r_g)$ & $ A_\phi$\\	\hline
        $ Bw $      & 11.6 & $ \sim r^{-3}~(n=-1) $ \\
        $ Bm $      & 13 & $ \sim r^{-3}~(n=-1) $ \\
        $ Bs $      & 14.8 & $ \sim r^{-2}~(n=0) $ \\
        $ Bvs $     & 15.5 & $ \sim r^{-1}~(n=1) $ \\
            \hline
    \end{tabular}

    \caption{
        A summary of the model parameters. The model names correspond to the initial magnetic field strength: very strong ($ Bvs $), strong ($ Bs $), moderate ($ Bm $), and weak ($ Bw $). The second column lists the corresponding magnetic field strength at $ r = 100\,r_g $, while the third column specifies the power-law index $ n $ used in the magnetic vector potential profile in Equation~\eqref{eq:A}.
        }
        \label{tab:models}

\end{table}

For the \textsc{h-amr} simulations, we use a spherical grid that is uniform in $(\log r, \theta, \varphi)$, where $(r, \theta, \varphi)$ are Kerr–Schild coordinates, and adopt a polytropic equation of state with a relativistic adiabatic index of $ \gamma = 4/3 $. We employ local adaptive time-stepping and one level of static mesh refinement in the radial range $10 < r/r_g < 100$, along with up to three levels of adaptive mesh refinement (AMR). The AMR criterion is based on the specific entropy $S$ of the fluid to ensure proper resolution of both the relativistic jet ($S > 10$) and the shocked jet region ($S > 0.1$). The radial grid extends from $ r_{\rm in} = 0.8\,r_g $ out to $ r_{\rm out} = 10^4\,r_g \approx 11\,R_{\rm He} $, where $r_g = G M_{\rm BH}/c^2 = 6 \times 10^6\,\cm $ is the gravitational radius of the BH. In all models except for $Bw$, the highest AMR level reaches a resolution of $4096 \times 2304 \times 4096$ cells in the $\hat{r}\text{--}\hat{\theta}\text{--}\hat{\phi}$ directions, respectively. To adequately resolve the magnetorotational instability (MRI) in the weak-field model $Bw$, we employ a higher maximum resolution of $6144 \times 6144 \times 8192$ cells. We verify that the resolution in all models is sufficient to resolve the fastest-growing MRI mode by ensuring that $Q \gtrsim 10$, where $Q$ is the ratio of the MRI wavelength to the proper cell size in the azimuthal direction, as required for reliable MRI resolution \citep{Hawley2011}.

\section{Magnetohydrodynamic evolution}\label{sec:evolution}

\begin{figure}[h]
  \centering
  \includegraphics[width=0.47\textwidth]{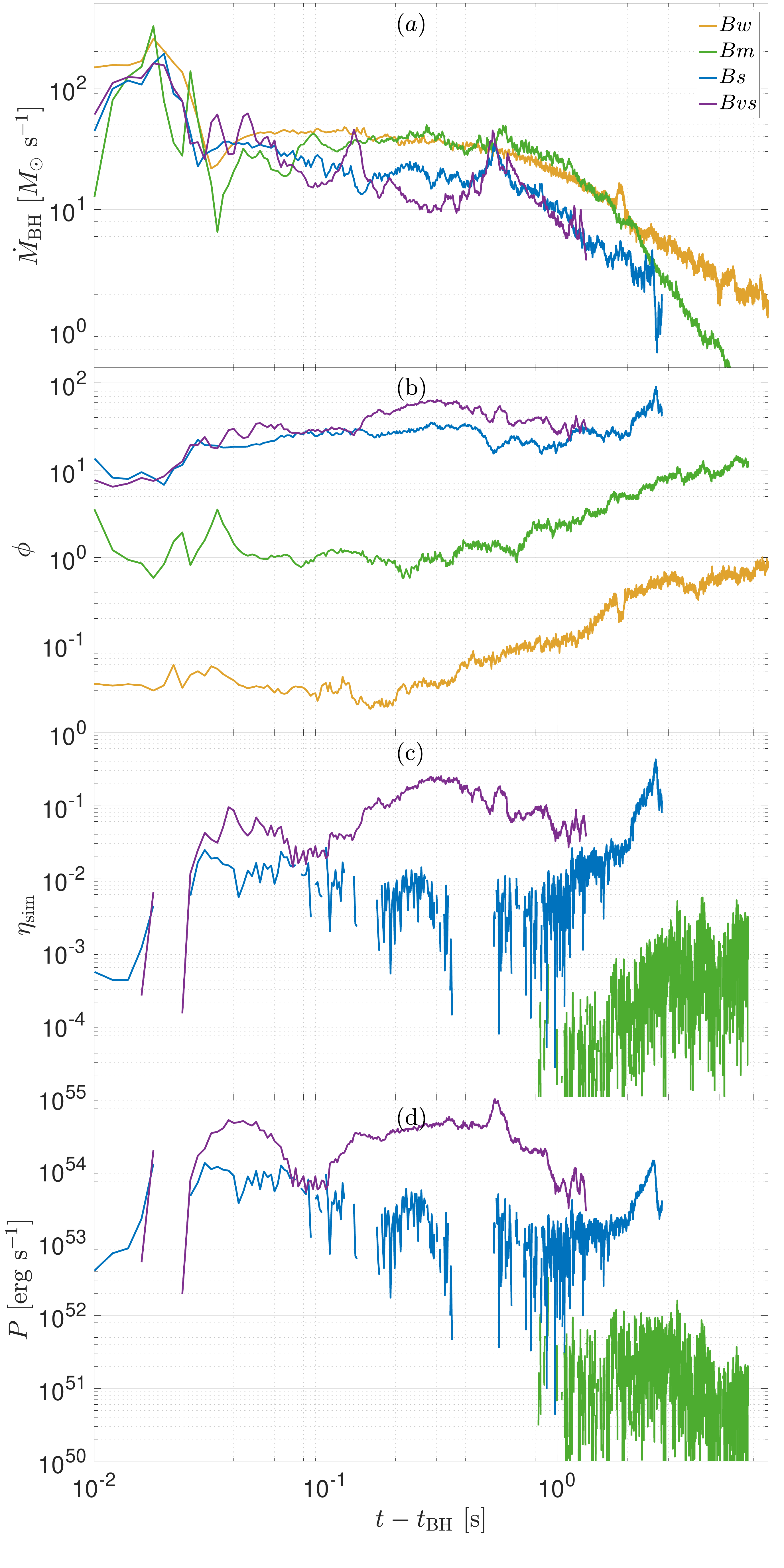}
  \caption{
Time evolution post-BH formation ($t-t_{\rm BH}$) of various quantities measured at the BH horizon (except for the mass accretion rate, which is calculated at $r = 5\,r_g$).
{\bf (a)} The mass accretion rate initially peaks at $\dmbh \gtrsim 100\,\msun\,\s^{-1}$ as the plasma free-falls onto the BH. This is followed by the formation of an accretion disk, which stabilizes the inflow until the bulk of the stellar material is accreted at $t \approx 1\,\s $, after which the accretion rate begins to decline.
{\bf (b)} The high dimensionless magnetic flux, $\phi$, indicates that rotational energy is extracted from the BH via magnetic torques in models $Bvs$ and $Bs$, leading to a non-negligible jet launching efficiency $\eta_{\rm sim} $ (panel {\bf (c)}). {\bf (d)} The resulting jets in models $Bvs$ and $Bs$ are extremely powerful, with luminosities $ P = \eta_{\rm sim}\dmbh c^2 \sim 10^{54}\,\erg\,\s^{-1}$, enabled by the combination of high mass accretion rates and strong magnetic fluxes threading the BH.
}
  \label{fig:evolution}
\end{figure}

Figure~\ref{fig:evolution}(a) shows that in the first few milliseconds post-BH formation, the dense core undergoes free-fall onto the BH, reaching extreme mass accretion rates of $ \dmbh \gtrsim 100\,\msun\,\s^{-1}$. This phase continues until an accretion disk forms, at which point the accretion rate stabilizes at $ \dmbh \sim 10\textsc{--}40\,\msun\,\s^{-1} $. This accretion rate persists until the BH begins to accrete the lower-density plasma that, at the time of BH formation, resided at $ r \approx 2\times 10^9\,\cm $ (indicated by the red line in Fig.~\ref{fig:density}), with a corresponding free-fall timescale of $ t_{\rm ff} \approx 1\,\s $. At that point, the mass accretion rate drops again and continues to decline as the remaining bound material is consumed by the BH.

Fig.~\ref{fig:evolution}(b) shows the evolution of the dimensionless magnetic flux,
\begin{equation}\label{eq:phi}
    \phi = \frac{\Phi}{\sqrt{\dmbh r_g^2c}}\,,
\end{equation}
which characterizes the dominance of magnetic flux $ \Phi $ on the BH horizon. In models $Bvs$ and $Bs$, $\phi$ reaches the saturation level $\phi_{\rm M} \approx 50$, indicating a magnetically arrested state \citep{Narayan2003,Tchekhovskoy2011}. This regime allows efficient extraction of rotational energy from the BH, enabling a steady launch of collimated relativistic outflows -- jets.

Fig.~\ref{fig:evolution}(c) presents the jet launching efficiency in the simulations,
\begin{equation}
    \eta_{\rm sim} = \frac{1}{\dmbh c}\int_{r_{\rm H}, \sigma>1}\sqrt{-g}(-T^r_t-\rho u^r)\mathrm{d}\theta \mathrm{d}\varphi\,,
\end{equation}
where the integration is calculated on the BH horizon $ r_{\rm H} = r_g\left(1+\sqrt{1-a^2}\right) $, and only for fluid elements with magnetization $ \sigma \equiv B^2/4\pi \rho c^2 > 1 $, $ g $ is the metric determinant, $ T^r_t $ denotes the radial energy flux density component of the mixed stress-energy tensor $ T $, and $ u^\mu $ is the four-velocity such that $ \rho u^r $ represents the radial mass-energy flux density. When the integrated momentum flux on the horizon is directed inward, $ \eta_{\rm num} < 0 $, whereas if the momentum flux carried by BH outflows overcomes the accretion power, the jet emerges from the ergosphere. Numerical estimates of the jet launching efficiency, $ \eta_{\rm num} $, show its dependency on $\phi$ and the BH spin $a$ \citep{Tchekhovskoy2011,Gottlieb2023e,Lowell2024},
\begin{equation}
    \eta_{\rm num} \approx \left(\frac{\phi}{\phi_{\rm M}}\right)^2\left(1.06a^4 + 0.4a^2\right)\,.
\end{equation}
For a spin of $a = 0.5$, the jets attain the theoretical maximum efficiency permitted by the extraction of BH rotation energy, $\eta_{\rm num, max} \approx 0.2$.

The combination of high accretion rates and elevated jet efficiency results in extremely powerful jets, as shown in Fig.~\ref{fig:evolution}(d), reaching $P = \eta_{\rm sim}\dmbh c^2 \sim 10^{54}\,\erg\,\s^{-1}$. We note that although our simulations do not self-consistently evolve the BH spin over time, the spin tends to increase when $\phi$ is low and decrease when $\phi$ is high \citep[e.g.,][]{Jacquemin-Ide2024, Issa_2025_spindown}. This variation in spin modifies the jet launching efficiency and luminosity relative to the fixed-spin values assumed in our calculations and shown in Fig.~\ref{fig:evolution}. We estimate the correction for the jet energy due to the BH spin evolution in \S\ref{sec:bh}, and discuss the implications of such electromagnetic transients in \S\ref{sec:summary}.

\begin{figure*}[]
  \centering
  \includegraphics[width=0.385\textwidth]{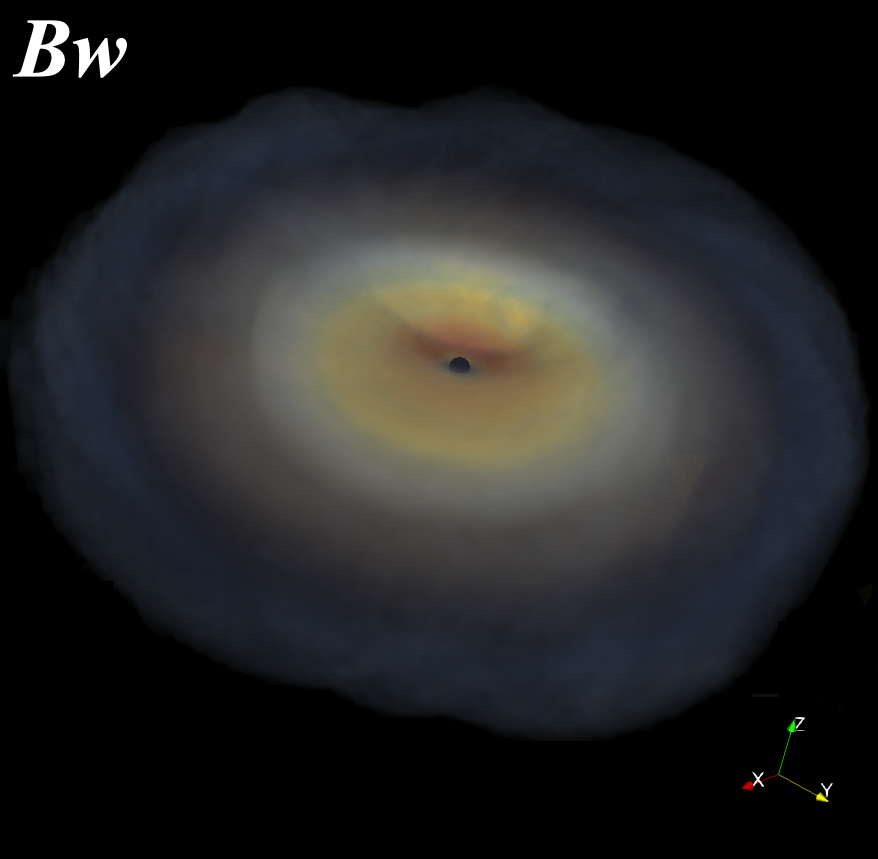}
  \includegraphics[width=0.31\textwidth]{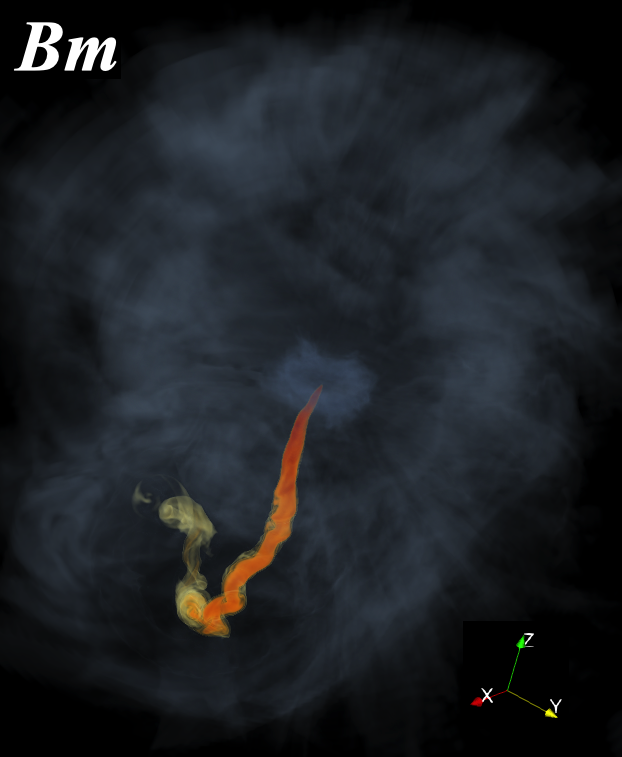}
  \includegraphics[width=0.273\textwidth]{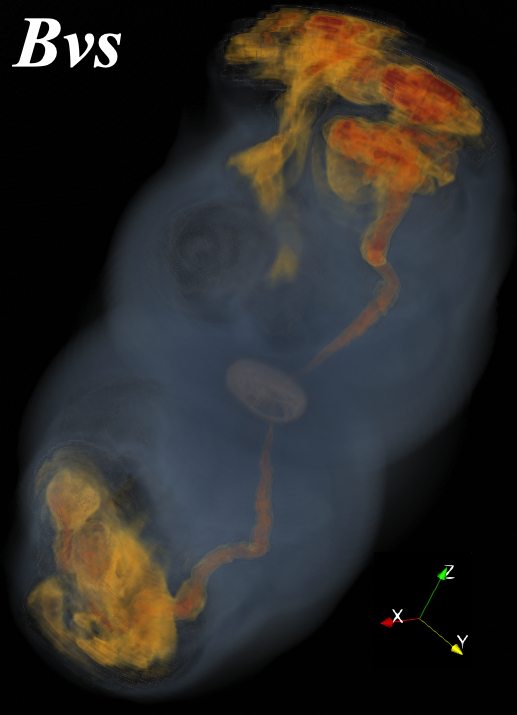}
  \caption{3D renderings of the late-time evolution in models $Bw$ (weak magnetic field) at $t-\tBH = 5.9\,\s$, $ Bm $ (moderate strength field) at $t-\tBH = 5.7\,\s$, and $Bvs$ (very strong field) at $t-\tBH = 1.3\,\s$. The $ \hat{z} $-axis aligns with the stellar rotation and BH spin. All models develop an accretion disk as a result of the star’s rotational profile. In model $Bw$, the rendering shows a rising mass density toward the BH, with no outflows emerging from the weakly magnetized accretion disk. By $t = 5.7\,\s$, the BH in this model has grown to $\mbh \approx 110\,M_{\odot}$, and we depict the BH horizon at $r_g = 1.65 \times 10^7\,\cm \sim 1\% $ of the disk radius. In contrast, model $ Bvs $ represents the opposite extreme, where relativistic outflows are electromagnetically launched from the BH and reach $ r \approx 4\,R_{\rm He} $. The accretion disk is shown in gray, the ultra-relativistic jet in red, and shocked jet (stellar) cocoon material in yellow (blue-gray). By $t = 1.3\,\s$, the outflows have unbound all plasma in the system, leaving the remaining gas either in the disk or in the cocoon. Model $ Bm $  represents an intermediate regime in which jets marginally emerge from the BH ergosphere, leading to intermittent jet launching. In the snapshot shown, only one jet is visible up to $ r \approx R_{\rm He} $, as the counter-jet is momentarily choked by the ram pressure of the infalling plasma.
}
  \label{fig:3d}
\end{figure*}

Figure~\ref{fig:3d} depicts 3D renderings of the system for three models: $ Bw, Bm, $ and $ Bvs $. Due to the compactness of the star relative to the BH radius, the accretion disk is noticeable in all models, extending to $ R_{\rm disk} \approx 0.1\,R_{\rm He} $. The evolution of the system is governed by the outflow power, which is primarily determined by the strength of the initial magnetic field.

In the weakly magnetized model $Bw$, the magnetic fields remain subdominant throughout the collapse, preventing both the BH and the disk from launching any outflows. As a result, Fig.~\ref{fig:3d} illustrates that the entire star is consumed by the BH, leaving behind a gradually shrinking accretion disk that persists for a few seconds. In model $Bm$, which features a moderate seed magnetic field, outflows are launched only after most of the stellar material has already been accreted and the mass accretion rate has dropped. Jets are launched once their Alfv\'{e}n velocity exceeds the free-fall velocity of the infalling gas, allowing magnetohydrodynamic waves to escape the BH ergosphere and form an outflow \citep{Gottlieb2023}. The declining $\dmbh$ increases the dimensionless magnetic flux to $\phi \approx 10$ (Equation~\eqref{eq:phi}), raising the Alfv\'{e}n velocity and enabling jet emergence \citep{Issa2025}. However, this magnetic flux is insufficient to sustain steady jet launching, resulting in unsteady, flickering outflows. Because the infalling gas is asymmetric relative to the BH polar axis, these outflows can also appear as one-sided jet episodes, as illustrated in Fig.~\ref{fig:3d}.

In the very strongly magnetized model $ Bvs $, Fig.~\ref{fig:3d} shows how the extreme magnetic fields launch relativistic jets early in the collapse, maintaining high jet-head velocities that allow them to punch through the star within a fraction of a second. The jets (red) shock the infalling plasma (yellow), generating an inflated, high-pressure cocoon (blue-gray) that helps collimate them. In doing so, the jets unbind the surrounding stellar material, preventing it from accreting onto the BH. Stochastic angular momentum transport from the cocoon into the disk induces a tilt in the accretion disk, leading to the observed wobbling motion of the jets \citep{Gottlieb2022c}.

\section{Black hole properties}\label{sec:bh}

To connect our simulations with the binary properties inferred for GW231123, we estimate the BH's final mass, spin, and kick velocity. Since our GRMHD simulations are performed in a fixed Kerr-Schild background and do not evolve the spacetime dynamically, we do not track the BH's growth or spin evolution self-consistently during the simulation. Instead, we compute these quantities in post-processing by integrating the mass and angular momentum accreted through the horizon to estimate the BH’s natal properties.

\begin{figure}[]
  \centering
\includegraphics[width=0.47\textwidth]{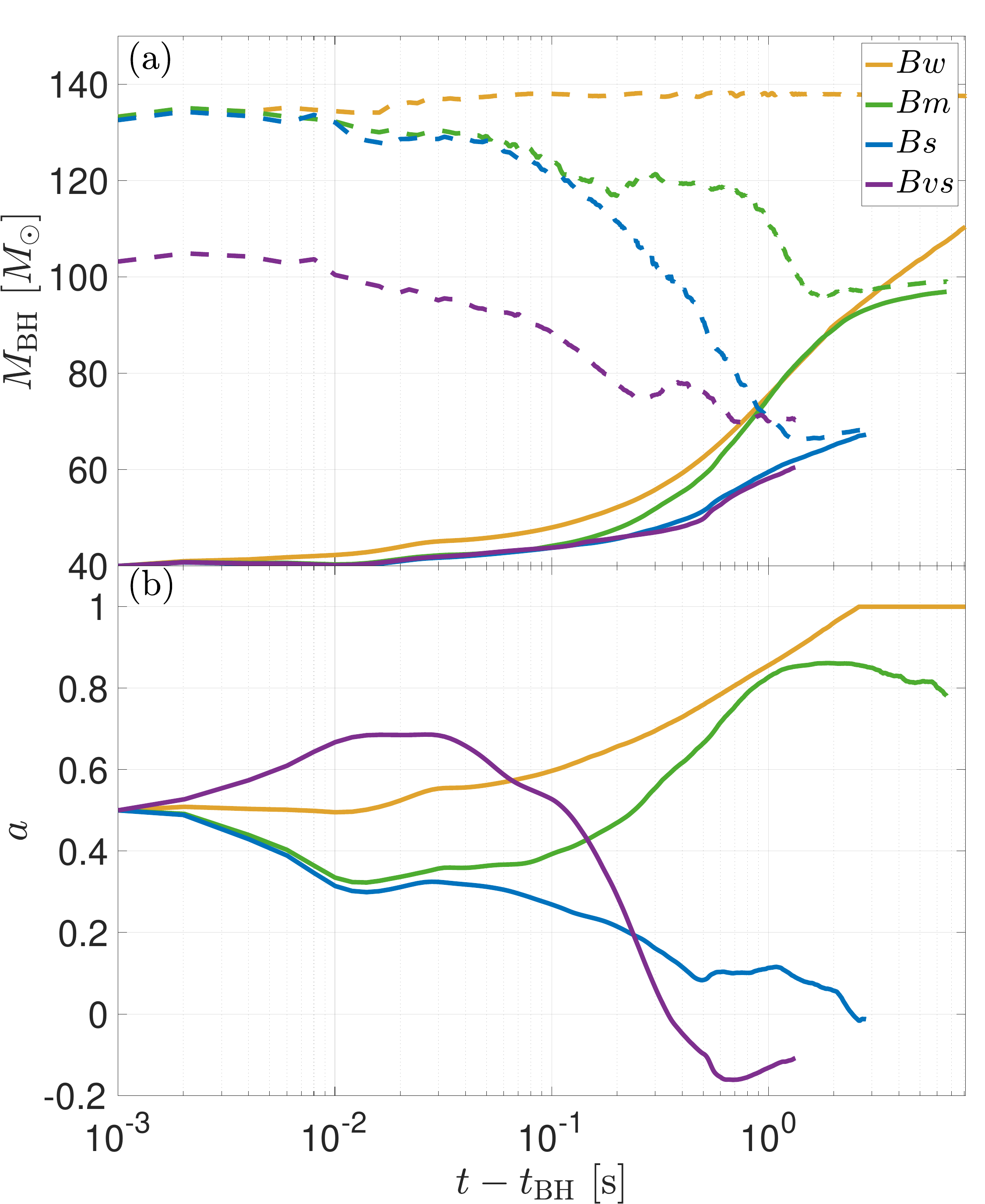}
\includegraphics[width=0.48\textwidth]{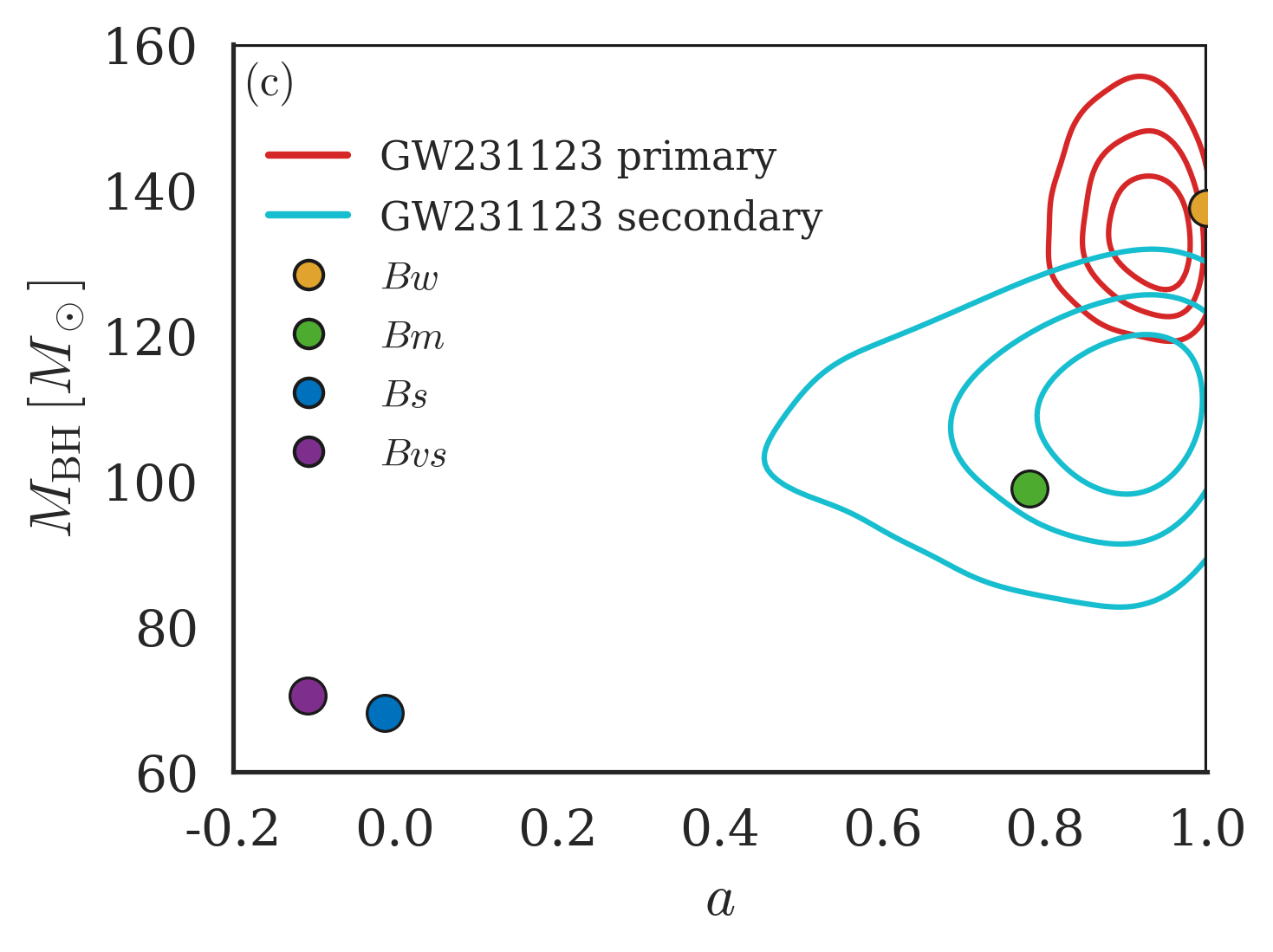}
  \caption{
{\bf (a)}: The time evolution of the BH mass (solid lines) and its asymptotic value, given by the sum of the BH mass and bound mass in the system (dashed lines), demonstrates that strong outflows halt accretion, unbind the surrounding plasma, and reduce the final BH mass. In all models, the BH mass converges to its asymptotic value by the end of the simulation.
{\bf (b)}: The time evolution of the BH spin. When no outflows are present (model $ Bw $), the BH spins up to the maximal value. In models with outflows, halted accretion and magnetic torques that extract rotational energy from the BH lead to spin-down. Strong outflows, as in models  $ Bs $ and $ Bvs $, drive the spin down toward an equilibrium value between accretion and extraction of angular momentum. In contrast, in model $ Bm $, where the outflows are launched later, the BH continues to spin up until jet launching, reaching $ a \approx 0.8 $.
{\bf (c)}: Final BH spin--mass (including all bound plasma at the end of our simulations) diagram. Colored circles indicate our models, while the contours are representative of the BHs inferred from GW231123 using the \textsc{NRSur7dq4} waveform \citep{Varma:2019csw}. The results show that a direct-horizon collapse binary with weak to moderate magnetic fields can reproduce the properties of the BHs observed in GW231123.
}
  \label{fig:BH}
\end{figure}

As outflows tap into the BH’s energy through accretion, the BH’s mass grows through the non-radiative component as
\begin{equation}
    \mbh(t) = \int_0^t \left[1-\eta_{\rm ev}\left(t'\right)\right]\dmbh\left(t'\right) \mathrm{d}t'\,,
\end{equation}
where
\begin{equation}
\eta_{\rm ev}(t) = \eta_{\rm sim}(t) \frac{\eta_{\rm num}\left(a\left(t\right)\right)}{\eta_{\rm num}(a=0.5)}\,
\end{equation}
denotes the efficiency corrected for an evolving BH spin.
Figure~\ref{fig:BH}(a) depicts the evolution of the BH mass (solid lines), along with its potential asymptotic value assuming it accretes all the remaining bound gas in the system (dashed lines), defined by the Bernoulli parameter criterion,
\begin{equation}
    -\left(1+\gamma \frac{u_g}{\rho c^2}+\sigma\right)u_t<1\,,
\end{equation}
where $ u_g $ is the internal energy of the plasma, and $ u_t $ is the covariant time-component of the four-velocity vector. A clear trend emerges across models: stronger magnetic fields drive stronger outflows, which unbind more stellar material. As a result, stars with stronger magnetic fields eject more mass, leaving less bound material available for BH growth. The BH in model $Bvs $ asymptotically reaches a final mass of $\mbh \approx (60-70)\,\msun$, whereas in the weak-field model $Bw$, the BH grows to nearly the full stellar mass, $\mbh \approx M_{\rm He}$. The intermediate model $ Bm $ gives rise to a BH within the upper mass gap. By the end of the simulations, the bound mass in the system converges to zero, indicating that the computed BH mass is a good approximation of its final value.

To compute the BH spin, we track the specific angular momentum, $ \ell $, accreted onto the BH. The spin evolves due to accretion as \citep{Gammie2004,Shapiro2005}
\begin{equation}
    \dot{a}_{\rm acc}(t) = \frac{\dmbh(t)}{\mbh(t)}\left[\frac{c\ell(t)}{G\mbh(t)}-2a(t)\left[1-\eta_{\rm ev}\left(t\right)\right]\right]\,,
    \end{equation}
where the first term describes spin-up from angular momentum accretion, while the second term represents spin-down due to the increase in BH mass.

When magnetic outflows efficiently extract angular momentum from the BH, its spin also evolves due to magnetic torques according to \citep{Moderski1996}
\begin{equation}
    \dot{a}_{\rm out}(t) = -\frac{\eta_{\rm ev}(t)c^3}{G\mbh(t)}\left[\frac{c^3}{G\mbh(t)k\Omega_{\rm H}(t)}-2a(t)\right]\,,
\end{equation}
where $ \Omega_{\rm H}(t) = a(t)c/2r_{\rm H}(t) $ is the angular frequency of the BH horizon, and $ k $ is a dimensionless parameter regulating the coupling between the BH spin and magnetic torques. We adopt the values of $ k $ appropriate for neutrino-cooled disks from \citet{Issa_2025_spindown}, which arise under the extreme mass accretion rates in these stars (Fig.~\ref{fig:evolution}(a)). The total spin change is thus
\begin{equation}
a(t) = \int_0^t \left[\dot{a}_{\rm acc}\left(t'\right) + \dot{a}_{\rm out}\left(t'\right)\right]\mathrm{d}t'\,.
\end{equation}

Fig.~\ref{fig:BH}(b) depicts the evolution of BH spin across the
different models. The assumed high angular momentum of the progenitor
star enables the BH to spin up through accretion. In model $Bw$, the
BH reaches maximal spin ($a = 1$) after roughly doubling its mass,
around $t \approx 2\,\s$ following its formation. All other models
exhibit outflows that prevent the BH from reaching $a = 1$ by
suppressing accretion and extracting its rotational energy. When
the outflows are launched late, the BH may still reach high spin values. This is seen in model
$ Bm $, where the spin stabilizes at $ a \approx 0.8 $ at
$ t \gtrsim 1\,\s $ coinciding with the onset of the outflows that
halt accretion and thereby freeze the spin evolution. This outcome is
robust against the absence of spin evolution in our simulations, as the higher jet power expected for $a > 0.5$ would suppress
accretion in the same manner.

In contrast, when magnetic fields become
dynamically important and the disk enters a magnetically arrested
state early on, as in models $ Bvs $ and $ Bs $, the BH spins down to an
equilibrium between accretion and magnetic extraction of angular
momentum. The interaction between the cocoon and the magnetically
arrested disk introduces stochastic angular momentum into the flow
(model $ Bvs $), disrupting the coherent inflow and triggering
accretion with opposite angular momentum. This, in turn, may also
activate the jittering-jet mechanism for exploding the star \citep{Papish2011,Papish2014}. As a result, after the jet has extracted the BH spin energy, subsequent accretion delivers angular momentum in the opposite direction, spinning it down to a retrograde value of $ a \approx -0.1 $. More accurately assessing the BH spin evolution will require dynamically evolving the BH properties and incorporating its feedback on the surrounding plasma.

To estimate the jet energy while accounting for spin evolution, we renormalize it using the expected spin-dependent efficiency,
\begin{equation}
E \approx \int_0^t P(t)\,\frac{\eta_{\rm ev}(t)}{\eta(t)}\,\mathrm{d}t\,.
\end{equation}
Incorporating our spin evolution calculation, the jet energy in model $Bvs$ decreases from $E \approx 3 \times 10^{54}\,\erg $ to $E\approx 6\times 10^{53}\,\erg$ due to spin-down. Similarly, in model $Bs$, the jet energy is reduced from $E \approx 7 \times 10^{53}\,\erg$ to $E\approx 2 \times 10^{52}\,\erg$. In contrast, the spin-up in model $Bm$ increases the jet energy from $E \approx 6 \times 10^{51}\,\erg$ to $E\approx 3\times 10^{52}\,\erg$.

Fig.~\ref{fig:BH}(c) shows the final BH spin--mass relation for all of our models (colored circles), including the contribution from bound plasma at the end of the simulation (dashed lines in Fig.~\ref{fig:BH}(a)). Our results reveal the expected correlation between BH spin and mass: strong magnetic fields lead to lower spin and mass due to suppressed accretion and efficient angular momentum extraction from the BH (models $ Bvs $ and $ Bs $), while weak magnetic fields allow the BH to accrete nearly all of the rotating plasma, producing high spins and masses. Our model with a moderate magnetic field, $Bm$, forms a BH with $a < 1 $ spin and mass within the upper mass gap, as some of the envelope is lost to outflows. Together with the weak-field model $Bw$, such a binary system can reproduce the component BHs inferred in GW231123 (contours).

Finally, the BH kick velocity can be estimated as $ v_k \sim \alpha M_{\rm out} v_{\rm out}/M_{\rm BH}, $ where $M_{\rm out}$ and $v_{\rm out}$ are the outflow mass and characteristic velocity, respectively, and $\alpha$ quantifies the degree of asymmetry in the outflows. For example, in model $Bm$, the outflow mass is $M_{\rm out} \approx 40\,M_\odot$, with material ejected at $v_{\rm out} \gtrsim 0.1\,c$. This implies that even a small asymmetry, $\alpha \approx 1\%$, may impart a BH kick of $ \sim 100\,{\rm km\,\s^{-1}} $. For a BH--helium core (after the secondary has lost its envelope), further Roche-lobe overflow (RLOF) is avoided for a semi-major axis obeying $ r \gtrsim r_{\rm min} \approx 2.6R_\star $ \citep{Paczynski1971,Eggleton1983}. Although most of the stellar mass is concentrated at $ r < R_{\rm He} \approx 5.6\times 10^9\,\cm $, the relevant radius for the onset of contact is the photosphere, $ R_\star \approx R_{\rm ph} \approx 5\times 10^{11}\,\cm $, because the extended envelope is in hydrostatic equilibrium and responds to the Roche potential.

To allow strong tidal-induced mixing consistent with CHE, we adopt an orbital period of $ P_{\rm orb} \sim 2\,{\rm d} $ \citep{deMink2009,Marchant2016,Song2016}, which corresponds to $ r \approx 3\times 10^{12}\,\cm $ and satisfies $ r > r_{\rm min} $. The orbital velocity is given by
\begin{equation}\label{eq:kick}
v_{\rm orb} \lesssim \sqrt{\frac{GM_{\rm He}}{2r}} \approx 7\times 10^2\,\left(\frac{M_{\rm He}}{150\,\msun}\frac{2\times 10^{12}\,\cm}{r}\right)^{1/2}\,\frac{\rm km}{\rm s}\,,
\end{equation}
which sets the characteristic scale for the BH kick velocity. Kicks with $ v_k \lesssim v_{\rm orb} $ are unlikely to unbind the system, whereas kicks comparable to or exceeding $ v_k $ can induce spin--orbit misalignment and, depending on their orientation, either disrupt the binary or drive it to a tighter orbit. A detailed numerical calculation of the kick dynamics will be presented in a follow-up paper (Li et al., in prep.).

\section{Discussion}\label{sec:summary}

GW231123 stands out as a landmark GW binary physics, marking the most massive binary BH merger observed to date. Its component masses and near-extremal spins place it well beyond the typical BH population seen by LIGO-Virgo-KAGRA, challenging standard formation scenarios such as isolated binary evolution, dynamical assembly, and hierarchical mergers. In conventional models, BHs forming from direct-horizon collapse above the pair-instability mass gap are expected to retain the full mass of their progenitors at core collapse, making it difficult to produce BHs within the gap. Here, we show explicitly for the first time that a massive, rotating, and magnetized progenitor star undergoing direct collapse to a BH can yield a remnant whose mass falls within the mass gap, supporting such progenitors as plausible sources of the GW231123 BHs.  A qualitatively similar explanation for generating the massive BH pair seen in GW190521 was proposed by \citet{Siegel+22}; however, this work adopted a semi-analytic model for the BH and ejecta properties that did not account for the effects of magnetic fields on the spin evolution.

Our end-to-end simulations span the evolution of a helium core, BH formation, and the subsequent evolution during stellar collapse. The helium core begins its life with a mass of $M_\star \approx 250\,\msun$ and is evolved as a single isolated star without rotation and without any hydrogen-rich envelope -- assuming that the latter does not influence significantly its evolution \citep[cf.][]{woosley:19, Renzo2020b, laplace:21}. After shedding mass due to stellar winds, our progenitor (now primarily composed of carbon and oxygen) encounters the pair-instability, but energy losses to photodisintegration prevent an explosion and lead to collapse with a total mass of $M \approx 150\,\msun$. About four seconds after the onset of collapse, a BH of mass $\mbh \approx 40\,\msun$ forms, initially accreting at a rate of $\dmbh \gtrsim 100\,\msun\,\s^{-1}$. Absent rotation and magnetic fields, the BH would eventually consume nearly the entire star, reaching a final mass of $\mbh \approx 150\,\msun$ \citep{Renzo2020b}.
A natural question is whether stars this massive actually exist, and if so, whether they occur in binaries. Two lines of evidence support their plausibility:

(i) Observations of the massive star population in 30 Doradus in the LMC suggest an upper initial stellar mass limit of at least $\gtrsim 200\,\msun $ \citep{schneider:18science}. At lower-than-LMC but finite metallicities, homogeneous evolution could yield progenitors resembling our models. While the existence of such extremely massive binaries remains uncertain, both the binary fraction and the average number of companions are observed to increase with stellar mass \citep[e.g.,][]{offner:23}. If these stars form in dense environments, as suggested by the most massive stars being concentrated in the central cluster R136 \citep{dekoter:97, crowther:10, crowther:16, brands:22}, dynamical processes are expected to assemble the most massive stars into tight binaries within $\lesssim 1$ Myr \citep{oh:16, fujii:11, banerjee:12, renzo:19:vfts682, stoop:24}.

(ii) The observation of long $\gamma$-ray bursts (GRBs) across redshifts \citep[e.g.,][]{lan:21}. Within the collapsar framework \citep{MacFadyen2001}, these require massive progenitors capable of retaining sufficient angular momentum to form accretion disks at collapse, even at finite metallicity. Binary evolution can facilitate this by both removing hydrogen-rich envelopes and supplying angular momentum, either through CHE \citep{maeder:00} or post-common-envelope tidal spin-up \citep[e.g.,][]{bavera:2020, bavera:22, sen:25}. Alternatively, late stellar mergers could augment both angular momentum \citep[e.g.,][]{chatzopoulos:20} and mass \citep[e.g.,][]{Kremer2020, Renzo2020, DiCarlo2020, ballone:23}, though mergers would also introduce fresh hydrogen-rich material and require a dynamical environment to yield GW progenitors.

Since the internal rotation and magnetic field structure of massive stars remain poorly constrained, we adopt a rotation profile from lower-mass collapsar progenitor models and explore the impact of various magnetic field configurations on the collapse dynamics and BH properties using a suite of GRMHD simulations. Our post-BH formation simulations reveal a clear correlation between BH mass and spin. In cases with weak magnetic fields, the BH accretes nearly the entire stellar angular momentum and mass, placing its final mass above the upper mass gap. In contrast, strong magnetic fields suppress accretion and extract angular momentum from the BH via magnetic torques, resulting in lower final masses and spins. This establishes a spin--mass correlation in massive collapsing stars, and more broadly, in rotating progenitors undergoing core collapse.

BHs formed in stellar collapse can inherit strong magnetic fields from a proto-magnetar \citep{Gottlieb2024b} or generate them via a disk dynamo \citep[e.g.,][]{Shibata2025}. Although direct-horizon-collapsing stars do not form a proto-neutron star, a dynamo action may still operate on the accretion timescale. \citet{Jacquemin_Ide_2024_dynamo} estimated that such a dynamo can amplify magnetic fields on a timescale of $t_{\rm dyn} \approx 10^4\,r_g/c \approx 2\,\s$, which coincides with the time at which the BH in our moderately magnetized model $Bm$ becomes magnetically dominated and begins launching outflows.

Remarkably, both the mass and spin of the remnant BH in model $ Bm $ are consistent with the properties of the secondary BH in GW231123, while those of our weak-field model $ Bw $ match the primary. If these two progenitors were paired in a binary \citep[see e.g.,][]{Popa2025}, and binary evolution processes did not significantly alter their dynamics, they would result in a GW231123-like merger. The mass ejected during the magnetic explosion can impart a non-negligible natal kick to the BH. In tight binaries, such kicks, depending on their orientation relative to the orbit, can determine the system’s fate: they may unbind the binary, induce spin--orbit misalignment, or widen the post-collapse orbit enough to permit long-delay mergers of low-metallicity progenitors at $ z < 1 $. We note that a spin–orbit misalignment accompanied by quasi-aligned spins, as proposed by \citet{Bartos2025}, is only expected to arise by chance in direct-horizon collapse binaries. Additional detections of similarly heavy mergers are needed to build stronger statistical evidence for spin alignment, which could serve as a discriminator among different formation models.

Although we observe a spin–mass correlation for this particular progenitor mass, other progenitor masses may yield different outcomes. For instance, a more massive and slowly spinning BH could result from a higher-mass progenitor, in which strong jet activity removes both mass and angular momentum, leading to a more massive but less rapidly spinning remnant. Alternatively, a slower rotation of the progenitor is expected to form a BH with higher natal mass but lower spin. Conversely, less-massive, rapidly spinning BHs, potentially such as those in GW190521 \citep{LIGO2020}, may arise either from a faster-spinning star that forms a BH with a lower natal mass, or a less massive progenitor that collapses just above the upper edge of the pair-instability mass gap, producing lower-mass, high-spin remnants through an evolution analogous to model $ Bm $.

The presence of magnetic fields in the progenitor star not only regulates the final BH mass and spin, but also has important implications for electromagnetic transients. The outflows launched by the BH carry substantial energy due to the high mass accretion rates in these events, making them extremely luminous sources. For our most energetic jet, the isotropic-equivalent power is $ P_{\rm iso} \sim 10^{56}\,\erg\,\s^{-1} $. Such powerful jets have not been observed, either due to their rarity stemming from both the scarcity of progenitor systems and the narrow beaming of relativistic outflows, or because the extreme magnetic flux required to produce them is difficult to realize in nature. Instead, we focus on model $Bm$, which produces a BH consistent with those inferred from GW231123, and we find that BH formation is accompanied by an electromagnetic transient with $P_{\rm iso} \sim 10^{54}\,\erg\,\s^{-1}$ sustained for several seconds. This energy output is comparable to that of the most energetic GRBs ever observed. Except for their short (rest-frame) duration relative to long GRBs, such transients may be observationally indistinguishable from the broader GRB population. A dedicated follow-up study is needed to investigate the electromagnetic signatures of these events in more detail.

The post-BH formation simulations do not incorporate radiation or neutrino transport, and we have not post-processed their output with a nuclear reaction network. Nevertheless, it is likely that a portion of the tens of solar masses of total ejecta will constitute radioactive elements, either in the form of $^{56}$Ni (e.g., \citealt{Dean&Fernandez24}) or, potentially, $r$-process nuclei \citep{Siegel+22,Issa2025}. Another prediction of the direct-horizon scenario for massive BH formation is therefore a supernova associated with each BH formation with a higher ejecta mass than is typical for ordinary collapsars from lower-mass stars. Such events could be observed following an extremely energetic GRB from the high-redshift universe (e.g., \citealt{Levan2024}).

\acknowledgements

We are grateful to Lucy Reading-Ikkanda/Simons Foundation for designing Figure~\ref{fig:sketch}. O.G. is supported by the Flatiron Research Fellowship. B.D.M. is supported by in part by the National Science Foundation (grant AST-2406637) and the Simons Foundation (grant 727700). D.I. is supported by Future Investigators in NASA Earth and Space Science and Technology (FINESST) award No. 80NSSC21K1851. M.R. acknowledges support from NASA (ATP: 80NSSC24K0932). The Flatiron Institute is supported by the Simons Foundation. This research used resources of the Argonne Leadership Computing Facility, a U.S. Department of Energy (DOE) Office of Science user facility at Argonne National Laboratory and is based on research supported by the U.S. DOE Office of Science-Advanced Scientific Computing Research Program, under Contract No. DE-AC02-06CH11357 (NeutronStarRemnants project). This research used resources of the National Energy Research Scientific Computing Center (NERSC), a Department of Energy User Facility using NERSC allocation m4603 (award NP-ERCAP0029085). Final computations in this work were run at facilities supported by the Scientific Computing Core at the Flatiron Institute, a division of the Simons Foundation.

\bibliography{refs}

\end{document}